\documentclass[num-refs]{nbdt-article}

\usepackage{siunitx}
\usepackage{ccfonts}
\usepackage{lmodern}
\usepackage{MnSymbol}
\usepackage{shortcut}
\usepackage{amsmath}
\usepackage{float}
\usepackage[normalem]{ulem}
\usepackage{graphicx}
\usepackage{caption}
\usepackage{subcaption}
\usepackage[]{bm}
\usepackage{pgfplotstable}
\pgfplotsset{compat=newest}
\usepgfplotslibrary{external}
\usetikzlibrary{arrows,calc,positioning,shapes.geometric,math}
\newcommand\numberthis{\addtocounter{equation}{1}\tag{\theequation}}

\papertype{Original Article}
\paperfield{Journal Section}

\title{Deep Recurrent Encoder: an end-to-end network to model magnetoencephalography at scale}

\abbrevs{MEG, magnetoencephalography; EEG, electroencephalography.}

\author[1,2\authfn{1}]{Omar Chehab}
\author[1\authfn{1}]{Alexandre D\'efossez}
\author[3]{Jean-Christophe Loiseau}
\author[2]{Alexandre Gramfort}
\author[1,4]{Jean-Remi King}

\contrib[\authfn{1}]{Equally contributing authors.}

\affil[1]{Facebook AI Research}
\affil[2]{Universit\'e Paris-Saclay, Inria, CEA, Palaiseau, France}
\affil[3]{ENSAM, Paris, France}
\affil[4]{\'Ecole normale sup\'erieure, PSL University, CNRS, Paris, France}

\corraddress{Jean-Remi King, Facebook AI Research}
\corremail{jeanremi@fb.com}

\fundinginfo{Alexandre Gramfort, Inria: ANR-20-CHIA-0016, and European Research Council Starting Grant SLAB ERC-StG-676943; Jean-Remi King, \'Ecole normale sup\'erieure, PSL University: ANR-17-EURE-0017 and the Fyssen Foundation.}

\runningauthor{Chehab et al.}

\begin{document}

\maketitle

\begin{abstract}
Understanding how the brain responds to sensory inputs from non-invasive brain recordings like magnetoencephalography (MEG)
can be particularly challenging:
(i) the high-dimensional dynamics of mass neuronal activity are notoriously difficult to model, (ii) signals can greatly vary across subjects and trials, and (iii) the relationship between these brain responses and the stimulus features is non-trivial.
These challenges have led the community to develop a variety of preprocessing and analytical (almost exclusively linear) methods, each designed to tackle one of these issues. Instead, we propose to address these challenges through a specific end-to-end deep learning architecture, trained to predict the MEG responses of multiple subjects at once. We successfully test this approach on a large cohort of MEG recordings acquired during a one-hour reading task. Our Deep Recurrent Encoder (DRE) reliably predicts MEG responses to words with a three-fold improvement over classic linear methods. We further describe a simple variable importance analysis to investigate the MEG representations learned by our model and recover the expected evoked responses to word length and word frequency. Lastly, we show that, contrary to linear encoders, our model captures modulations of the brain response in relation to baseline fluctuations in the alpha frequency band. The quantitative improvement of the present deep learning approach paves the way to a better characterization of the complex dynamics of brain activity from large MEG datasets.

\end{abstract}

\section{Introduction}
\label{sec:intro}

A major goal of cognitive neuroscience consists of identifying how the brain responds to distinct experimental conditions.
While descriptive statistics and statistical tests are classically used to analyze neural data \citep{maris2007eegstats}, this approach is not suited to predict how the brain should react to new conditions. The resulting models of the brain can thus be particularly challenging to compare. By contrast, predictive encoding models \citep{naselaris2011encoding,king2018encoding} can be directly trained to predict brain responses to various experimental conditions, and compared on their ability to accurately predict novel conditions. For example, encoding models allow the estimation of integration constants in the brain \citep{shivangi2020lstmtimescales,mesgarani2019}, the hierarchical organization of visual \citep{yamins2016using} and speech processing~\citep{caucheteux2022brains,millet2021inductive}. Beyond MEG, predictive models have enabled automatic segmentation  \citep{Petreska2011} and dynamical system identification~\citep{duncker2019fixedpoint,maheswaranathan2019fixedpoint}. 
In functional Magnetic Resonance Imaging, predictive encoding models are starting to emulate complex neural processing~\citep{marblestone2016dlneuro} and are a step towards discovering new phenomena \citep{seeliger2018convolutional,seeliger2021end}.
Yet, this general objective of developing encoding models faces three major challenges when working with non-invasive and time-resolved signals collected by magneto- and electro-encephalography (M/EEG).

\paragraph{Challenge 1: rich response dynamics}
M/EEG signals are known and promoted for their excellent temporal resolution. While this ability to measure cognitive processes at a millisecond time-scale offers unique opportunities for fine chronometry of neural responses in humans, it also makes such signals notoriously difficult to analyze. For example, brain responses to audio streams overlap in time making their identification difficult. To address this issue in the context of encoding models, it is standard to employ a Temporal Receptive Field (TRF) model~~\citep{STRF,smith2015regression1,smith2015regression2,holdgraf-etal:17,crosse-etal:16,goncalves-etal:14, martin-etal:14,lalor2009neural,Ding5728,sassenhagen:19}\footnote{also referred to as Finite Impulse Response (FIR) analysis in fMRI~\citep{POLINE2012871}, and Distributed Lag modeling in statistics~\citep{cromwell1994multivariate}}. TRF models are commonly designed to predict neural responses to exogenous stimulation by fitting a linear regression model with a fixed time-lag window of past sensory stimuli. By doing so, the predictions derived from TRFs are only influenced by stimuli descriptors, enabling them to modulate their response based on previous brain activity. Consequently, unless the basal activity from previous time points is introduced as an exogenous feature, TRF cannot learn to 
capture neuronal adaptation responses \citep{adaptation}, nor can it learn to vary an evoked response as a function of the pre-stimulus alpha power ~\citep{attentionalpha, vanrullen2016perceptual}.

\paragraph{Challenge 2: inter-trial and inter-subject variability}
Neuronal recordings in general, and M/EEG in particular, can be extremely variable across trials and subjects~\citep{baillet:17, king2018encoding, iturrate2014latency, vanrullen2016perceptual,huth2016natural}.
To reduce the nuisance factors behind these variations such as eye blinks, head movements, cardiac, and face muscle activity which corrupt MEG recordings, it is common to make use of multiple sessions and subjects within a study. For example, several methods based on spatial filtering \citep{icacorrection, sspcorrection, barachant2011multiclass,sabbagh-etal:20,parra2005} or "hyper-alignment" use linear models such as canonical correlation analysis (CCA), partial least square regression (PLS), multi-view ICA and back-to-back regressions (B2B)~\citep{de2019multiway,xu2012regularized,Lim2017,king2020back,bazeille2020sharedresponse,richard2020ica} to isolate the brain responses shared across trials and/or individuals. However, these denoising techniques can also remove relevant signals. For example, \citet{vanrullen2016perceptual} have repeatedly shown that evoked responses to sensory input can be modulated by pre-stimulus alpha activity in a predictable way. Averaging trials, or filtering out this variability during preprocessing would therefore prevent the identification of such phenomenon.

\paragraph{Challenge 3: identifying the relationship between brain responses and stimulus features}  
A large part of cognitive neuroscience aims to identify \emph{how} the brain responds to stimulus features. For example, are V1 neurons tuned to respond to luminance, contrast, oriented lines, or faces? When and where does this elicit a response? To tackle this issue, it is common to present many stimuli to the subject, and fit a general linear model (GLM) to predict brain responses given a set of hypothetical features \citep{naselaris2011encoding}.
This approach can be limited, as GLMs only reveal brain responses to features predetermined by the analyst~\citep{poldrack2008guidelines,POLINE2012871,Kording2018,ivanova2021simple} and understanding interactions between features often requires explicitly modeling these interactions (e.g. as cross-terms to form a quadratic polynomial) and feeding them to a linear regression \citep{gareth2013statlearning}.

Here, we propose to simultaneously address these three core challenges with a unique end-to-end ``Deep Recurrent Encoding" (DRE) neural network trained to robustly predict brain responses from both (i) past MEG activity and (ii) current experimental conditions. We test DRE on 99 subjects recorded with MEG during a one-hour long reading task, and show that our model (1) better predicts MEG responses than standard models, (2) efficiently captures inter-trial and inter-subject variability, and (3) identifies feature-specific responses as well as interactions between basal activity and these evoked responses.


\section{Materials and Method}
\label{sec:materials_and_method}

This section presents, with consistent and self-contained mathematical notations, a 
methodological progression from linear to nonlinear encoding models of neural dynamics as observed with MEG. It also discusses the statistical and computational benefits of recurrent models, as well as the novel methodological ideas proposed with the DRE model.

\subsection{Problem formalization}
\label{ssec:problem_and_notations}

In the case of MEG, the measured magnetic fields $x$
reflect a tiny subset of brain dynamics $h$ --- specifically a partial and macroscopic summation of the synaptic input to cortical pyramidal cells. Given the physics of electromagnetic fields propagation, it is standard to assume that these neuronal magnetic fields have a linear, stationary, and instantaneous relationship with the magnetic fields measured via MEG sensors \citep{hamalainen1993magnetoencephalography}. We refer to this ``readout operator'' as $C$, a matrix which is subject-specific because it depends on the location of pyramidal neurons in the cortex and thus on the anatomy of each subject. Furthermore, the brain dynamics governed by a function $f$ evolve according to their past and to external stimuli $u$ \citep{wilson1972excitatory}. In sum, we can formulate the problem as follows:
\begin{align*}
    \begin{cases}
    x_{\mathrm{current}} & = C h_{\mathrm{current}} \\
    h_{\mathrm{current}} & = f(h_{\mathrm{past}}, u_{\mathrm{current}})
    \end{cases}
\end{align*}

\subsection{Operational Objective}

Here, we aim to parameterize $f$ with $\theta$, and subsequently learn $\theta$ and $C$ to obtain a statistical (as opposed to biologically constrained as in \citep{friston_dynamical_causal_modeling}) generative model of observable brain activity that accurately predicts MEG activity $\hat x \in \mathbb{R}^{d_x}$ given an initial state and a series of past stimuli.

\paragraph{Notations} We denote by $u_t \in \mathbb{R}^{d_u}$ the stimulus with $d_u$ encoded features at time $t$, $x_t \in \mathbb{R}^{d_x}$ the MEG recording with $d_x$ sensors, $\hat{x}_t$ its estimation, and by $h_t \in \mathbb{R}^{d_h}$ the underlying brain activity. Because the true underlying brain dynamics are never known, $h$ will always refer to a model estimate. To facilitate the parametrization of $f$, it is common in the modeling of dynamical systems to explicitly create a ``memory buffer" by concatenating successive lags. We adopt the bold notation $\bm{h}_{t-1:t-\tau_h} := \left[h_{t-1}, ..., h_{t-\tau_h}\right] \in \mathbb{R}^{d_h \tau_h}$ for flattened concatenation of $\tau_h \in \mathbb{N}$ time-lagged vectors. With these notations, the dynamical models considered in this paper are described as:
\begin{align}
    \begin{cases}
    x_{t} & = C h_{t} \\
    h_{t} & = f_\theta(\bm{h}_{t-1:t-\tau_h}, \bm{u}_{t:t-\tau_u})
    \end{cases}
 \label{eq:genericdyn}
\end{align}
where 
\begin{itemize}
    \item $f: \mathbb{R}^{d_h \tau_h + d_u (\tau_u + 1)} \rightarrow\mathbb{R}^{d_h}$ governs brain dynamics given the preceding brain states and external stimuli
    \item $C \in \mathbb{R}^{d_h \times d_x}$ is a linear, stationary, instantaneous, and subject-specific observability operator that makes a subset of the underlying brain dynamics observable to the MEG sensors.
\end{itemize}

\subsection{Models}
\label{ssec:models}

\paragraph{Temporal Receptive Field (TRF)}
Temporal receptive fields (TRF)~\citep{STRF} are arguably the most common model for predicting neural time series in response to exogeneous stimulation. The TRF equation is that of control-driven linear dynamics: 
\begin{equation}
     h_{t} = f_{\theta}(\bm{h}_{t-1:t-\tau_h}, \bm{u}_{t:t-\tau_u}) = B\bm{u}_{t:t-\tau_u}  \enspace , \numberthis 
\end{equation}
where $B \in \mathbb{R}^{d_h \times d_u . (\tau_u + 1)}$ is the convolution kernel that maps the stimuli to the brain response and $\theta = \{B\}$. By definition, the TRF kernel encodes the input-output properties of the system, namely, its characteristic time scale, its memory, and thus its ability to sustain an input over time. A computational drawback is that the TRF kernel size scales linearly with the duration of the neural response to the stimulus. For example, a dampened oscillation evoked by the stimulus could last one hundred time samples ($\tau_u = 99$) and would require $B \in \mathbb{R}^{d_h  \times 100 d_u}$ to reach 100 steps in the past, even though oscillatory dynamics can be compactly written as a second-order differential equation expressing $h_t$ in terms of only two of its own past states $(h_{t-1}, h_{t-2})$\footnote{A sine wave can be produced by a simple linear auto-regressive (AR) model of order 2}. Emulating this, we will introduce a recurrent component to the TRF model to tackle the issue of dimensionality.

\paragraph{Recurrent Temporal Receptive Field (RTRF)} A Recurrent Temporal Receptive Field (RTRF) is a linear auto-regressive model. The RTRF with exogenous input can model time-series from its own past (e.g., past brain activity) \textit{and} from exogenous stimuli. Unrolling the recurrence reveals that current brain activity can be expressed in terms of past activity. This corresponds to recurrent dynamics with control:
\begin{align*}
    h_t & = f_{\theta}(\bm{h}_{t-1:t-\tau_h}, \bm{u}_{t:t-\tau_u}) = A\bm{h}_{t-1:t-\tau_h} + B\bm{u}_{t:t-\tau_u}  \enspace , \label{eq:linearautoregression} \numberthis
\end{align*}
where the matrix \( A \in \mathbb{R}^{d_h \times (d_h . \tau_h)} \) encodes the recurrent dynamics of the system and $\theta = \{A, B\}$. 

The dependency of $h_t$ on $h_{t-1}$ in \eqref{eq:linearautoregression} means we need to
unroll the expression of $h_{t-1}$ in order to compute $h_t$. However, it has been shown that linear models
perform poorly in this case, as terms of the form $A^t$ ($A$ to the power $t$) will appear, with either exponentially exploding
or vanishing eigenvalues. This rules out optimization with first order methods due to the poor conditioning
of the problem~\citep{bottou2018optimization}, or using a closed-form solution.
To circumvent unrolling the expression of $h_{t-1}$, we need to obtain it from what is measured at time $t-1$. This however assumes the existence of an inverse relationship from $x_{t-1}$ to $h_{t-1}$, which we assume here to be linear by using the pseudo inverse of $C$: $h_{t-1} = C^{\dagger} x_{t-1}$. As a result, $h_t$ and $x_t$ are identifiable to one another, and \eqref{eq:linearautoregression} can be solved in closed form as a regular linear system~\citep{billings2013identificationbook}. Initializing the RTRF dynamics with the pre-stimulus data can be written as:
\begin{align*}
    h_{t} & = C^{\dagger} x_{t} \qquad \forall t \in \{0, ..., \tau_h - 1\} \enspace , \label{eq:linearautoreg_init} \numberthis
\end{align*}
where $\tau_h$ is chosen to match the pre-stimulus duration $\tau$.

Though the recurrent component of the RTRF is able to reduce the receptive field $\tau_u$ of TRF, it is nevertheless constrained to maintain a `sufficiently big' receptive field $\tau_h$ to initialize over $\tau_h$ steps. The following model, DRE, will avoid this issue, and will also not require that $h_t$ and $x_t$ are identifiable via linear inversion.

\paragraph{Deep Recurrent Encoder (DRE)}
DRE is an architecture based on the Long-Short-Term-Memory (LSTM) computational block \citep{hochreiter1997long}. It is useful to think of the LSTM as a ``black-box nonlinear dynamical model'', which composes the RTRF building block with nonlinearities and a memory module which reduces the need for receptive fields, so that $\tau_h = 1$ and $\tau_u = 0$. It is employed here to capture nonlinear dynamics evoked by a stimulus. A single LSTM layer can be formulated as~\citep{hochreiter1997long}:
\begin{align}
\begin{cases}
 h_{t} & = f_{\theta}(\bm{h}_{t-1:t-\tau_h}, \bm{u}_{t:t-\tau_u}) = o_{t} \odot \tanh (m_{t}) \\
 m_{t} & = d_{t} \odot m_{t-1} + i_{t} \odot \tilde{m}_{t} \\
 \tilde{m}_{t} & = \tanh ( Ah_{t-1} + Bu_{t} )
\end{cases} \enspace ,
\numberthis \label{eq:LSTM}
\end{align}
where the $\tanh$ nonlinearity is applied element-wise, $\odot$ is the Hadamard (element-wise) product, and $(d_t, i_t, o_t) \in ( \mathbb{R}^{d_m} )^3$ are data-dependent vectors with values between 0 and 1 modeled as forget (or drop) input and output gates, respectively. 
The memory module $m_t \in \mathbb{R}^{d_m}$ thus interpolates between a ``past term" $m_{t-1} \in \mathbb{R}^{d_m}$ and a ``prediction term" $\tilde{m}_{t} \in \mathbb{R}^{d_m}$, taking $h_t$ as input. The ``prediction term" (See \eqref{eq:LSTM} last equation) resembles that of the previous RTRF model except that it is here composed with a tanh nonlinearity which conveniently normalizes the signals.

Again, the dependency of $h_t$ on $h_{t-1}$ in \eqref{eq:linearautoregression} meant that we needed to
unroll the expression of $h_{t-1}$ to compute $h_t$. While this is numerically unstable for the RTRF, the LSTM is designed such that $h_t$ and its gradient
are stable even for large values of $t$. As a result, $h_t$ and $x_t$ do not need to be identifiable to one another. In other words, contrary to RTRF, the LSTM allows $h_t$ to represent a hidden state containing potentially more information than its corresponding observation $x_t$.

We now motivate three modifications made to the standard LSTM.

First, we help it recognize when (not) to sustain a signal, by augmenting the control $u_t$ with a mask embedding $p_t \in \{0, 1\}$ indicating whether the provided MEG signal generates the current brain response (i.e. 1 before word onset and 0 thereafter).
Second, we automatically learn to align subjects with a dedicated subject embedding layer. Indeed, a shortcoming of standard brain encoding analyses is that they are commonly performed on each subject separately. However, this implies that one cannot exploit potential similarities across subjects. Here, we adapt the LSTM in the spirit of~\citet{SING2018} so that a \textit{single} model is able to leverage information across multiple subjects. We do this by augmenting the control $u_t$ with a ``subject embedding'' $s \in \mathbb{R}^{d_s}$, that is learned for each subject. Note that this amounts to learning a matrix in $\mathbb{R}^{d_s \times n_s}$ that is applied to the one-hot-encoding of the subject number. In order words, each subject has a vectorized representation that is one column of the embedding matrix. Setting $d_s < n_s$ allows us to use the same LSTM block to model subject-wise variability, and to train across subjects simultaneously while leveraging similarities across subjects.

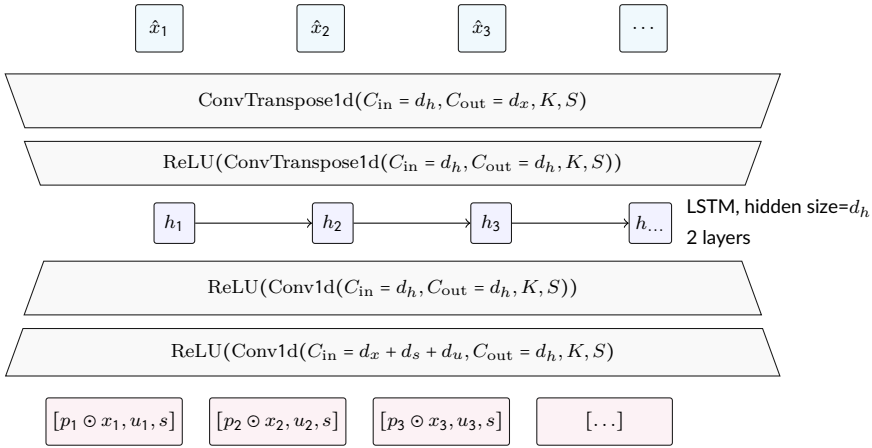
\begin{figure}[!ht]
    \centering
    \def\pscale{.9}

\begin{tikzpicture}[
    every node/.style={scale=\pscale},
    conv/.style={shape=trapezium,
        trapezium angle=70, draw, inner xsep=0pt, inner ysep=2pt,
        draw=black!90,fill=gray!5},
    deconv/.style={shape=trapezium,
        trapezium angle=-70, draw, inner xsep=0pt, inner ysep=2pt,
        draw=black!90,fill=gray!5},
    rnn/.style={rounded corners=1pt,rectangle,draw=black!90,
        fill=blue!5,minimum width=0.6cm, minimum height=0.6cm},
    inp/.style={rounded corners=1pt,rectangle,draw=black!90,
        fill=purple!5,minimum width=2cm, minimum height=0.7cm},
    outp/.style={rounded corners=1pt,rectangle,draw=black!90,
        fill=cyan!5,minimum width=0.7cm, minimum height=0.7cm},
    skip/.style={line width=0.2mm, ->},
]
    \def\yshift{0.6em}
    \def\base{12cm}
    \def\dec{0.55cm}

    \node (e1) [conv, minimum width=\base - \dec, anchor=south] at (0, 0)
        {$\mathrm{ReLU}(\mathrm{Conv1d}(C_{\mathrm{in}}=d_x + d_s + d_u, C_{\mathrm{out}}=d_h, K, S)$};
    \node (e2) [conv, minimum width=\base - 2*\dec, anchor=south] at
        ($(e1.north) + (0, \yshift)$)
        {$\mathrm{ReLU}(\mathrm{Conv1d}(C_{\mathrm{in}}=d_h, C_{\mathrm{out}}=d_h, K, S))$};
        
    \node (ls0) [rnn] at ($(e2.north) + (-0.35 * \base + 2 * \dec + 0.2cm,0.5cm)$) {$h_1$};
    \foreach \k/\text in {1/$h_2$,2/$h_3$,3/$h_{\ldots}$} {
        \tikzmath{
            int \prev;
            \prev=\k - 1;
        }
        \node (ls\k) [rnn,anchor=west] at ($(ls\prev.east) + (1.55cm, 0)$) {\text};
        \draw [->] (ls\prev) -- (ls\k);
    }
    \node (ls3) [anchor=west] at ($(ls3.east) + (0.1cm, 0)$) [align=left] {LSTM, hidden size=$d_h$\\ 2 layers};

    \node (d2) [deconv, minimum width=\base - 2*\dec, anchor=south] 
        at ($(e2.north) + (0, 1.cm)$) {$\mathrm{ReLU}(\mathrm{ConvTranspose1d}(C_{\mathrm{in}}=d_h, C_{\mathrm{out}}=d_h, K, S))$};
    \node (d1) [deconv, minimum width=\base - \dec, anchor=south] at
        ($(d2.north) + (0, \yshift)$) {$\mathrm{ConvTranspose1d}(C_{\mathrm{in}}=d_h, C_{\mathrm{out}}=d_x, K, S)$};

    \node (input1) [inp] at ($(e1.south) + (-0.4 * \base + 2 * \dec,-0.6cm)$) {$[p_1 \odot x_1, u_1, s]$};
    \node (input2) [inp,anchor=west] at ($(input1.east) + (0.35cm, 0)$) {$[p_2 \odot x_2, u_2, s]$};
    \node (input3) [inp,anchor=west] at ($(input2.east) + (0.35cm, 0)$) {$[p_3  \odot x_3, u_3, s]$};
    \node (input4) [inp,anchor=west] at ($(input3.east) + (0.35cm, 0)$) {$[\ldots]$};
    \node (output1) [outp] at ($(d1.north) + (-0.35 * \base + 2 * \dec,0.6cm)$) {$\hat{x}_1$};
    \node (output2) [outp,anchor=west] at ($(output1.east) + (1.5cm, 0)$) {$\hat{x}_2$};
    \node (output3) [outp,anchor=west] at ($(output2.east) + (1.5cm, 0)$) {$\hat{x}_3$};
    \node (output4) [outp,anchor=west] at ($(output3.east) + (1.5cm, 0)$) {$\ldots$};
\end{tikzpicture}
    \caption{
    Representation of the Deep Recurrent Encoder (DRE) model used to predict MEG activity. The masked MEG $p_t \odot x_t$ enters the network from the bottom,
    along with the control representation $u_t$ and the subject embedding $s$.
    The encoder transforms the input with convolutions and ReLU nonlinearities.
    Then, the LSTM models the sequence of hidden states $\bm{h}_t$, which are converted back to the MEG activity estimate $\hat{x}_t$.
    $\mathrm{Conv1d}(C_{\mathrm{in}}, C_{\mathrm{out}}, K, S)$ represents a convolution over time with $C_{\mathrm{in}}$ input channels, $C_{\mathrm{out}}$ output channels,
    a kernel size $K$, and a stride $S$. Similarly, $\mathrm{ConvTransposed1d}(C_{\mathrm{in}}, C_{\mathrm{out}}, K, S)$ represents a transposed convolution over time.
}
    \label{fig:model}
\end{figure}

Third, for comparability purposes, RTRF and LSTM should access the same pre-stimulus MEG information $\bm{x}_{\tau:1}$.
Incorporating the initial MEG, before word onset, is done by augmenting the control with $p_t \odot x_t$. The extended control reads: $\tilde{u}_t = \left[u_t, s, p_t, p_t \odot x_t\right]$, and 
the LSTM with augmented control $\tilde{u}_t$ finally reads:
\begin{align*}
    h_t & = f_{\theta}(\bm{h}_{t-1:t-\tau_h}, \bm{\tilde{u}}_{t:t-\tau_u})
    = \mathrm{LSTM}_{\theta}(h_{t-1}, \tilde{u}_t)
    = \mathrm{LSTM}_{\theta}(
        \mathrm{LSTM}_{\theta}(h_{t-2}, \tilde{u}_{t-1}), \tilde{u}_t
    )
    \enspace . \label{eq:LSTM_augmented} \numberthis
\end{align*}
In practice, to maximize expressivity, two modified LSTM blocks are stacked on top of one another (Figure~\ref{fig:model}).

Having introduced a nonlinear dynamical system for the brain response $h_t$, we can also extend the model \eqref{eq:genericdyn} by challenging the linear instantaneous mixing from the brain response $h_t$ to the measurements $x_t$. Introducing two new nonlinear functions $d$ and $e$, respectively parametrized by $\theta_2$ and $\theta_3$, a more general model formally reads:
\begin{align}
    \begin{cases}
    \bm{x}_{t:t-\tau_x+1} & = d_{\theta_2}(h_t) \\
    h_{t} & = f_{\theta_1} (\bm{h}_{t-1:t-\tau_h},
                            e_{\theta_3}(\bm{\tilde{u}}_{t:t-\tau_u}))
    \end{cases} \enspace ,
 \label{eq:genericdyn_nonlin}
\end{align}
where $\tau_x$ allows us to capture a small temporal window of data around $x_t$, and $\tau_u$
is taken to be much larger than $\tau_x$.
Indeed \eqref{eq:genericdyn_nonlin} corresponds to \eqref{eq:genericdyn} if one sets $\tau_x=1$ and $d_{\theta_2}(h_t) = C h_t$, as well as
$e_{\theta_3}(\bm{\tilde{u}}_{t:t-\tau_u}) = \bm{u}_{t:t-\tau_u}$.
In more intuitive terms, the DRE model generalizes the linear instantaneous measurement of the previous models with a ``convolutional autoencoder" \citep{Masci2011StackedCA}. The $e$ (encoder) function is formed by convolutions and the $d$ (decoder) function uses transposed convolutions, where both functions are two layers deep (Figure \ref{fig:model})
\footnote{While ``encoding'' typically means outputting the MEG 
with respect to the neuroscience literature, we use ``encoder'' and ``decoder'' in the context
of deep learning auto-encoders \citep{hinton2006reducing} in this paragraph.}.

In practice, we use a kernel size $K=4$ for the convolutions. This impacts the receptive field of the network and the parameter $\tau_x$. Equation \eqref{eq:genericdyn_nonlin} implies that the number of time samples in $h$ and $x$ are the same. However, a strong benefit of the convolutional auto-encoder is to perform a reduction of the number of time steps by using a stride $S$ larger than 1. By using a stride of 2, one reduces the temporal dimension by 2. Indeed it boils down to taking every other time sample from the output of the convolved time series. Given that the LSTM module is by nature sequential, this reduces the number of time steps it has to consider when learning, which accelerates both training and evaluation.
Further, there is evidence that LSTMs can only pass information over a limited number of time steps \citep{koutnik2014clockwork}.
In practice, we use $d_h$ output channels for the convolutional encoder.

In summary, our DRE model generalizes TRF and RTRF models by using nonlinearities both in the dynamics of the brain response $h_t$ and in its measurement $x_t$~\eqref{eq:genericdyn}. It is done respectively with LSTM cells and a convolutional auto-encoder. Importantly, the DRE is equipped with a subject embedding allowing us to learn a joint model for the group of subjects.

\subsection{Optimization losses}
\label{ssec:losses}
The dynamics for the above three models (TRF, RTRF, DRE) are given by different expressions of $f_{\theta}$ as well as the mappings between $x$ and $h$ via $C$ for TRF and RTRF, or $c$ and $e$ for DRE.

At test time, the models aim to accurately forecast MEG data from initial steps combined with subsequent stimuli. Consequently, one should train the models in the same setup. This boils down to minimizing a ``multi-step-ahead" $\ell_2$ prediction error:
\begin{align*}
\begin{cases}
    \minimize_{\theta_1, \theta_2, \theta_3} & \sum_t \| x_t - \hat{x}_t \|_2^2 \\
    \mathrm{s.t.}
    \quad & \bm{x}_{t:t-\tau_x+1} = d_{\theta_2}(h_t) \\
          & h_{t} = f_{\theta_1} (\bm{h}_{t-1:t-\tau_h},
            e_{\theta_3}(\bm{\tilde{u}}_{t:t-\tau_u}) )
\end{cases}
\end{align*}

This ``multi-step-ahead" minimization requires unrolling the recurrent expression of $h_t$ over the preceding time steps, which the LSTM-based DRE model is able to do~\citep{teacher_forcing_nlp}. 
The DRE model takes as input the observed MEG at the beginning of the sequence, and must predict the future MEG measurements using the (augmented) stimulus $\tilde{u}_t$. 
Note that the mapping to and from the latent space, $e_{\theta_3}$ and $d_{\theta_2}$, are learned \textit{jointly} with the dynamical operator $f_{\theta_1}$.
Furthermore, the DRE has reduced temporal receptive fields, thus the computational load is lightened and allows for a low or high-dimensional latent space. 

Given that the RTRF model is linear and can suffer from numerical instabilities (see above), it is trained with the ``one-step-ahead'' version of the predictive loss with squared $\ell_2$ regularization:
\begin{align*}
\begin{cases}
    \minimize_{\theta} & \sum_t \| x_t - \hat{x}_t \|_2^2 + \lambda \| \theta \|_2^2 \\
    \mathrm{s.t.}\quad & \hat{x}_t = C h_{t} \\
                       & h_t = f_{\theta}(\bm{h}_{t-1:t-\tau_h}, \bm{u}_{t:t-\tau_u}) \\
                       & \bm{h}_{t-1:t-\tau_h} = C^{\dagger} \bm{x}_{t-1:t-\tau_h}
\end{cases}
\end{align*}

TRF models are also trained with this ``one-step-ahead'' loss. 
As mentioned above, the linear models (TRF) require a larger receptive field than the nonlinear DRE. Large receptive fields induce a computational burden, because each time lag comes with a spatial dimension of size $d_h$ or $d_u$. To tackle this issue, $C$ is chosen to reduce this spatial dimension. In practice, we choose to learn $C$ \textit{separately} from the dynamics to simplify the training procedure of the linear models. Given a participant, we fit a Principal Component Analysis (PCA) with 40 components on their averaged (evoked) MEG data: the resulting PCA map is taken to be the matrix $C \in \mathbb{R}^{d_x \times 40}$. The resulting latent space will thus explain most of the variance of the original recording. Indeed, when training the TRF model on all 270 MEG sensors with no hidden state ($6.4 \pm 0.22\%$, MEAN and SEM across subjects) or using a 40-component PCA ($6.43 \pm 0.17$\%), we obtained similar performances. The pseudo-inverse $C^{\dagger}$ required to compute the previous latent state $h_{t-1}$ is also obtained from the PCA model. Note that dimensionality reduction via linear demixing is a standard preprocessing step in MEG analysis \citep{sspcorrection, Jung-etal:01, de2018decoding}. 

\subsection{Model Evaluation}
\label{ssec:model_evaluation}

\paragraph{Evaluation Method}
Following seminal works (\emph{e.g.} by \citet{kay2008,vangerven2017fmri}),
models are evaluated using the Pearson Correlation R (between -1 and 1) between the model prediction $\hat x$ and the true MEG signals $x$ for each channel and each time sample (after the initial state) independently\footnote{Figure~\ref{fig:explainedvar} in the Appendix~\ref{sec:appendix} reports the same evaluations, using a different metric: the explained variance quantified by the coefficient of determination $R^2$ of the brain response by the model predictions.}.
When comparing the overall performance of the models, we average over all time steps after the stimulus onset, and over all MEG sensors for each subject independently.

\paragraph{Feature Importance}
To investigate what a model actually learns, we use Permutation Feature Importance \citep{Breiman} which measures the drop in prediction performance when the $j^{th}$ input feature $u^j$ is shuffled:
\begin{align*}
    \Delta R_j=R - R_{j} \enspace , \numberthis
\end{align*}
By tracking $\Delta R$ over time and across MEG channels, we can locate in time and space the contribution of a particular feature (e.g. word length) to the brain response.

\paragraph{Experiment} 
The model weights are optimized with the training and validation sets, while the penalization $\lambda$ for the linear models (TRF and RTRF) is optimized with a grid search over five values distributed logarithmically between \( 10^{-3} \) and \( 10^3 \). Training of the DRE is performed with ADAM~\citep{ADAM} using a learning rate of $10^{-4}$ and PyTorch's default parameters~\citep{pytorch} for the running averages of the gradient and its square. The training is stopped when the error on the validation set increases. In practice, the DRE and the DRE-PCA were trained over approximately 20 and 80 epochs, respectively.

\paragraph{Statistics}
Each subject score is obtained using the model prediction on held-out trials, using the learned subject embeddings as the ``alignment function", similar to \citet{chen2015sharedresponse,zhang2018sharedresponse,haxby2011sharedresponse,bazeille2020sharedresponse,richard2020ica}.
To test the reliability of our effects (e.g. prediction performance, feature importance, model comparison), we assess confidence intervals and p-values across subjects using a non-parametric Wilcoxon rank test across subjects. When applicable, we correct these estimates for multiple comparisons using a false discovery rate (FDR) across time samples and channels. Note that subjects can be treated as independent observations to derive meaningful p-values since the statistics are based on held-out data independent from the training set.

\paragraph{Noise Ceiling}
Noise ceilings are typically estimated using batches of repeated conditions~\citep{sahani2002noiseceiling,kay2013compressive,vangerven2017fmri}, to evaluate the maximal amount of explainable variance. This involves multiple presentations of the same stimulus characterized by a given feature set.
In our case, however, sentences are presented only once to each subject. Further, we cannot control one of the variables input to our encoding models: namely, the baseline brain activity.

\label{sec:experiment}

\subsection{Data}
\label{ssec:data}
\paragraph{Experimental design}
We analyze 99 subjects from the Mother Of Unification Studies (MOUS) dataset \citep{MousMEG} who performed a one-hour reading task while being recorded with a 273-channel CTF MEG scanner. The task consisted of reading approximately 2,700 words flashed on a screen in rapid series. Words were presented sequentially in groups of 10 to 15, with a mean inter-stimulus interval of 0.78 seconds (min: 300ms, max: 1,400ms). Sequences were either random word lists or actual sentences (50\% each). For this study, both conditions were used. However, this study does not investigate the differences obtained across these two conditions. Out of the original 102 subjects, 3 were discarded from the study because we could not reliably parse their stimulus channels.

\paragraph{Stimulus preprocessing}
We focus on four well-known features associated with reading, namely word length (i.e., the number of letters in a word), word frequency in natural language (as derived by the wordfreq Python package \citep{robyn_speer_2018_1443582}, and measured on a logarithmic scale), and a binary indicator for the first and last words of the sequence. At a given time $t$, each stimulus $u_t \in  \mathbb{R}^{4}$ is therefore encoded with four values, fed to the models as a square function that is non-zero for the duration of the stimulus. Each feature is standardized to have zero mean and unit variance. Word length is expected to elicit an early (from t=100 ms) visual response in posterior MEG sensors, whereas word frequency is expected to elicit a late (from t=400 ms) left-lateralized response in anterior sensors. In the present task, word length and word frequency are correlated R=-0.48. 

\paragraph{MEG Preprocessing} As we are primarily interested in evoked responses \citep{tallon1999oscillatory}, we band-pass filtered between 1 and 30\,Hz and downsampled the data to 120\,Hz using the MNE software \citep{MNEPython} with default settings: i.e. a FIR filter with a Hamming window, a lower transition bandwidth of 1\,Hz with -6\,dB attenuation at 0.50 Hz and a 7.50\,Hz upper transition bandwidth with an attenuation of -6\,dB at 33.75\,Hz.

To limit the interference of large artefacts on model training, we use Scikit-learn's RobustScaler with default settings ~\citep{scikit-learn} to normalize each sensor using the 25\textsuperscript{th} and 75\textsuperscript{th} quantiles. Following this step, most of the MEG signals will have a scale around 1. Since we observed a few large scale outliers, we chose to reject any segment of 3 seconds that contains amplitudes higher than 16 in absolute value (fewer than 0.8\% of the time samples).

These continuous data are then segmented between 0.5\,s before and 2\,s after word onset, yielding a three-dimensional MEG tensor per subject: words, sensors, and time samples relative to word onset. For each subject, we form a training, validation, and test set using respectively 70\%, 10\%, and 20\% of these segments, ensuring that two segments from different sets do not originate from the same word sequence to avoid information leakage. This corresponds to 191K, 27K, and 53K  segments used for the train, validation, and test sets, respectively. Each segment has a spatial dimension of 273 sensors and a temporal dimension of 300 time points (2.5\,s sampled at 120\,Hz).

For clarity, some figures use global field power (GFP) to summarize effects over time. GFP refers to the standard deviation across MEG channels of an average evoked response.

\subsection{Model training}
\label{ssec:model_training}

\paragraph{Model hyper parameters}
We compare the three models introduced in Section~\ref{ssec:models} over $n_s = 99$ subjects. 
For the TRF, we use a lag on the control of $\tau_u = 250$ time steps (about 2\,s). This corresponds to the duration of the signal after the stimulus onset.
For the RTRF, we use $\tau_u = \tau_h = 40$ time steps. This lag is close to the minimum inter-word duration of 300\,ms, and corresponds
to the part of the initial MEG (i.e. 333\,ms out of 500\,ms, at 120\,Hz) that is passed to the model to predict the 2.5\,s MEG sequence during the evaluation.

For the DRE model, we use a subject embedding of dimension $d_s = 16$, a latent state of dimension $d_h = 512$, a kernel size $K=4$, and a stride $S = 2$. The subject embeddings are initialized as Gaussian vectors with zero mean and unit variance, while the weights
of the convolutional layers and the LSTM are initialized using the default ``Kaiming'' initialization~\citep{kaiming_init}.
Like its linear counterpart, the DRE is given the first 333\,ms of the MEG signal to predict the complete 2.5\,s of a training sample. 

\paragraph{Ablation study}
To investigate the importance of the different components of the DRE model, we implement an ablation study by fitting the model with all but one of its components. To this end, we compare the DRE to i) the DRE without using the 333\,ms of pre-stimulus initial MEG data (DRE NO-INIT), ii) the DRE trained in the 40-dimensional PCA space used for the linear models (DRE PCA), iii) the DRE devoid of a subject embedding (DRE NO-SUBJECT), and to iv) the DRE devoid of the convolutional auto-encoder (DRE NO-CONV).

\subsection{Code availability}
The code developed for the present study is available at~\href{https://github.com/facebookresearch/deepmeg-recurrent-encoder}{https://github.com/facebookresearch/deepmeg-recurrent-encoder}.

\section{Results}
\label{sec:results}

\begin{figure}[hbt!]
\centering
\includegraphics[width=0.8\columnwidth]{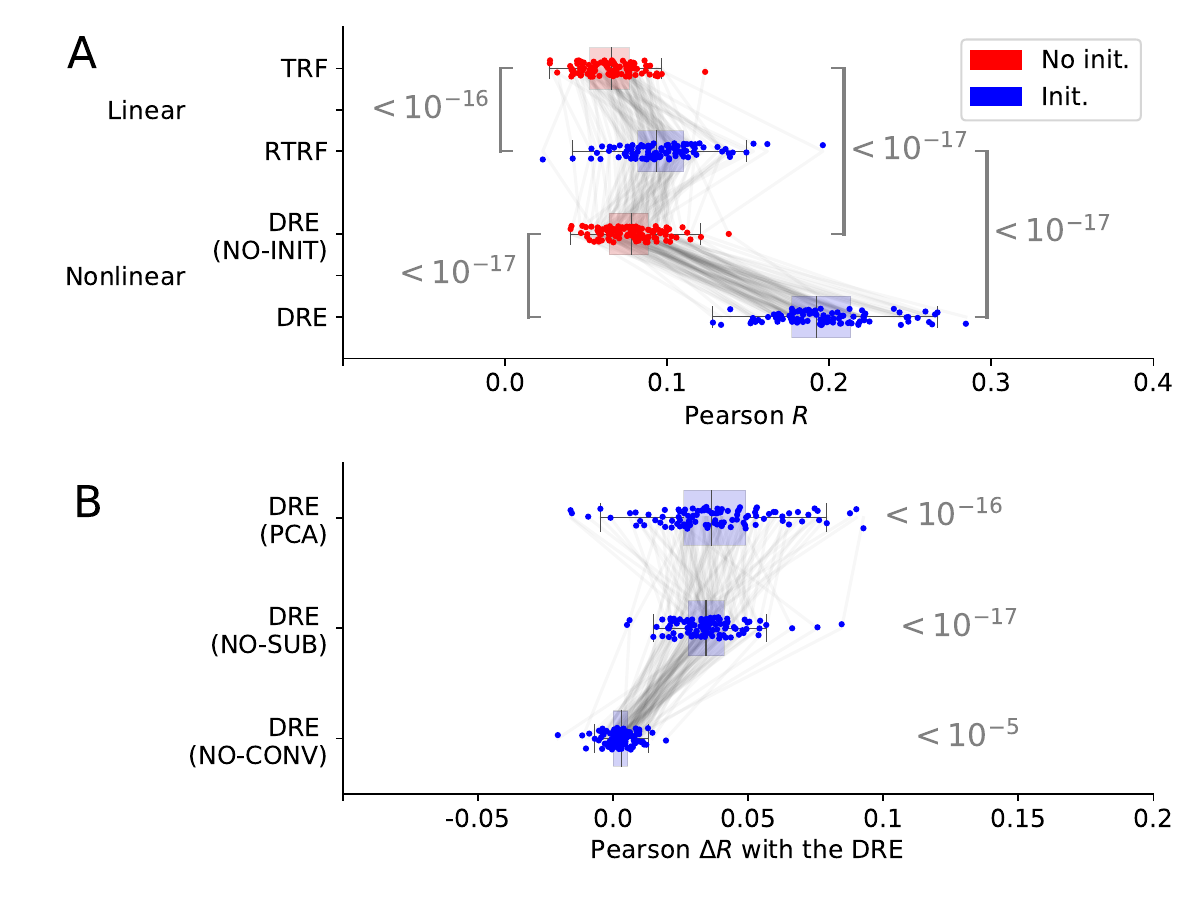}
\caption{
\textit{A: Model Comparison}. Predictive performance of different models over 99 subjects (dots) (chance=0). Boxplots indicate median and interquartile range, while gray lines follow each subject. The gray brackets indicate the p-value obtained with a pair-wise Wilcoxon comparison across subjects. Initialization and nonlinearity increase predictive performance.
\textit{B: Ablation Study of our model (DRE)}. Pearson R mean obtained by removing a component from (Convolutional Auto-encoder or Subject-embedding) or adding an element (PCA embedding composed with the Convolutional Auto-encoder) to the DRE architecture and retraining the model, and the resulting p-values obtained by comparing the resulting scores to those of the normal DRE. Convolutions marginally impact performance of the DRE (they are essentially used for speed), but training over all MEG sensors (vs. PCA space) and using subject embeddings designed for realignment are responsible for an increase of approximately 4\%.}
\label{fig:modelcomparison}
\end{figure}
We first evaluate the DRE's ability to predict brain responses to written words presented in rapid serial visual presentation and  measured with MEG, and compare these brain predictions to those of linear encoding models (TRF, RTRF). 
Then, we show with ablation experiments which elements of the DRE help address the challenges of rich dynamics, inter-subject, and inter-trial variability. 
Finally, we show how feature importance helps address the third challenge introduced above, namely: identifying the relationship between brain responses and stimulus features.
\begin{figure}[hbt!]
\centering
\includegraphics[width=0.8\columnwidth]{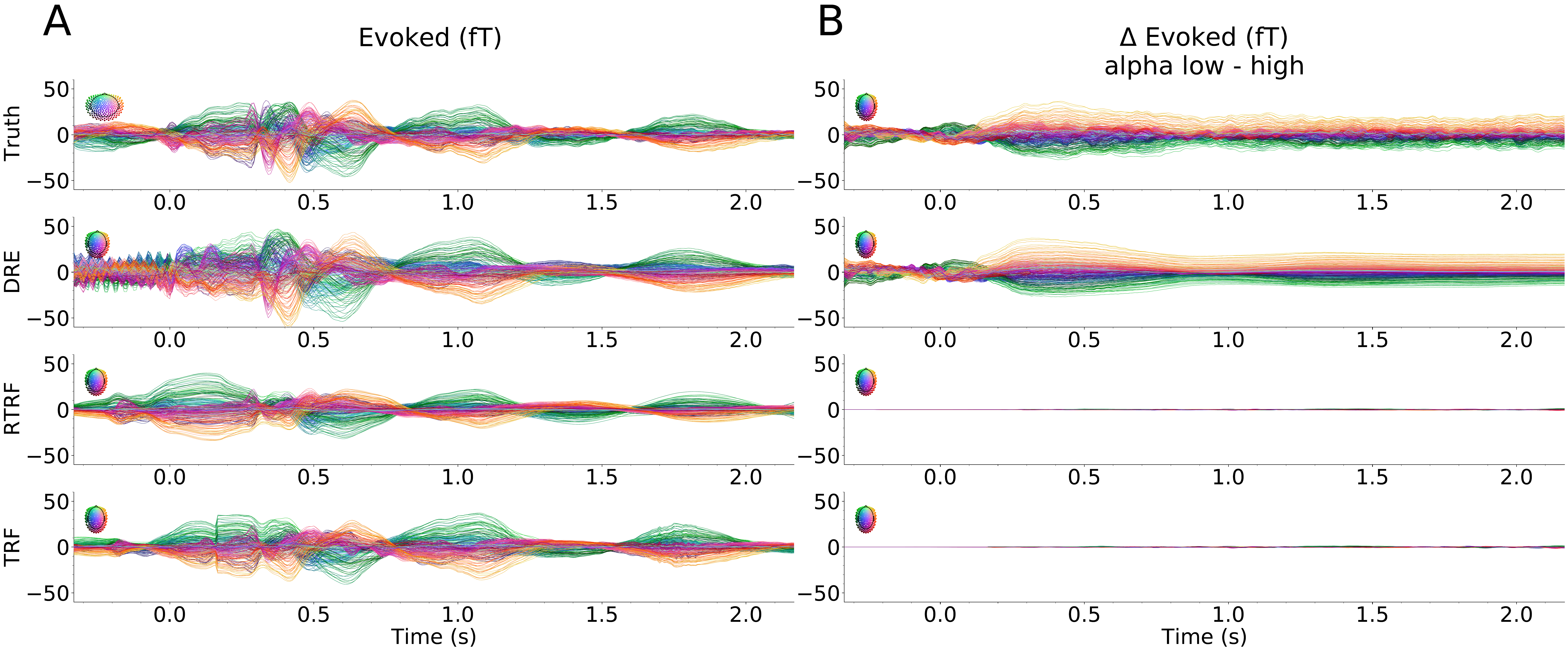}
\includegraphics[width=0.8\columnwidth]{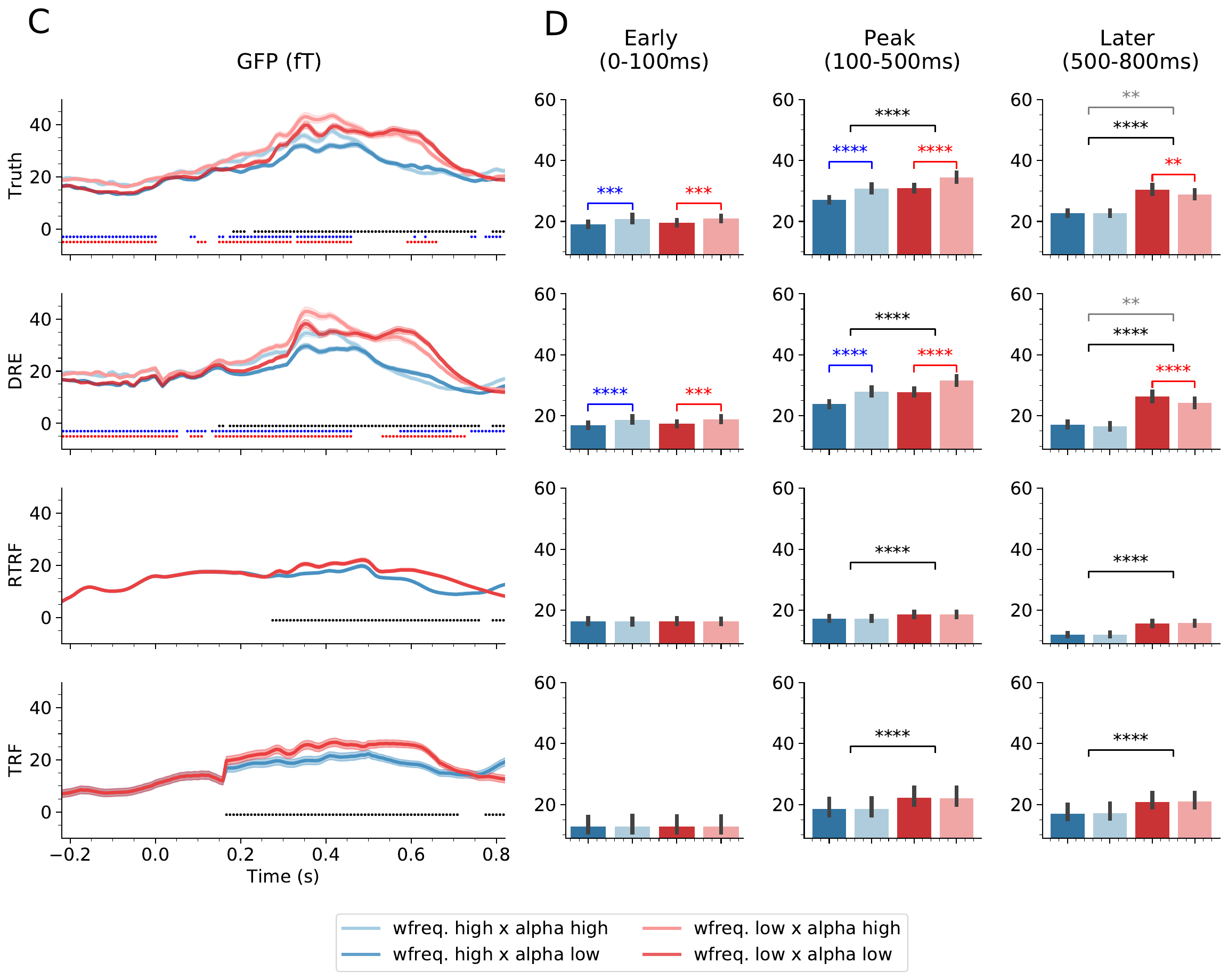}
    \caption{
    \textbf{A:}
    Brain response obtained for all three models (TRF, RTRF, DRE) and the actual data (Truth), averaged over the test set for all 99 subjects. Qualitatively, all models learn the mean MEG response, although with a variable precision.
    \textbf{B:}
    Difference of average brain response between trials with high \emph{vs} low pre-stimulus alpha power (8-13\,Hz). This analysis shows that DRE modulates the predicted brain response as a function of pre-stimulus alpha power.
    \textbf{C:}
    Global Field Power (GFP) of the brain response, as a function of pre-stimulus alpha power and word frequency.
    Qualitatively, all models capture the stimulus-modulation of the brain response, but only the DRE captures the alpha-modulation, which varies with the latency.
    \textbf{D:}
    Global Field Power (GFP) of the brain response averaged before the peak (0-100ms), during the ramp up of the peak (100-500ms) and after the peak of the evoked response (500-800ms). The number of stars illustrates the level of significance (**: $10^{-2}$, ***: $10^{-3}$, or ****: $10^{-4}$).
    The marginal pre-stimulus alpha effect (light vs. dark) is followed by the marginal word frequency effect (black) and then an interaction (grey) between pre-stimulus alpha and word frequency (red vs. blue). Only DRE captures the same effects as in the actual data.
    }
\label{fig:alpha_effect}
\end{figure}
\paragraph{Modeling rich MEG dynamics: model comparison.} DRE outperforms the baseline TRF and RTRF models with up to a three-fold improvement (Figure \ref{fig:modelcomparison}).
To provide a fair comparison between the models, we also compare TRF to a NO-INIT DRE, i.e. to a DRE architecture that ignores the pre-stimulus MEG activity. The results show that DRE NO-INIT consistently outperforms TRF (mean correlation score $R=0.077$ on average across all subjects, time samples, and all channels; standard error of the mean across subjects: $\pm$ 0.002 for DRE NO-INIT vs. $R=0.064 \pm 0.002$ for TRF). This difference is strongly significant ($p<10^{-17}$) 
under a Wilcoxon test across subjects. 
Similarly, DRE (R=$0.20 \pm 0.003$) significantly ($p<10^{-17}$) outperforms RTRF (R=$0.10 \pm 0.003$), when both of these models are given as input the pre-stimulus MEG activity. To verify that this gain is not trivially accounted for by the limited dimensionality of RTRF (trained with 40-dimensional Principal Components because of computational limitations), we trained DRE with the \textit{same} PCA-reduced data as RTRF. The results confirm that DRE obtains a higher performance (R=$0.16 \pm 0.003$, $p<10^{-16}$) than RTRF. Overall, these results suggest that DRE better models the rich M/EEG dynamics than linear models.

\paragraph{Subject embeddings efficiently capture inter-individual variability}
To evaluate the importance of the subject-embedding layer in capturing inter-individual variability, we trained the DRE without a subject embedding layer (DRE NO-SUB). 
The comparison between DRE and DRE NO-SUB reveals a clear difference ($\Delta R=0.038$, $p<10^{-17}$). This result shows that the subject embedding layer efficiently re-aligns subjects' brain responses to model the dynamics specific to each -- or shared across -- subject(s).

\paragraph{Recurrence efficiently captures inter-trial variability.}
Brain responses to sensory input are known to vary with ongoing brain activity\citep{vanrullen2016perceptual,haegens2018rhythmic}.
Recurrent models (RTRF, DRE) are thus well suited to capture such phenomenon: initialized with 333\,ms of pre-stimulus MEG, they can use basal brain activity to modulate the post-stimulus MEG predictions. Our results confirm this prediction: 
TRF is outperformed by RTRF ($0.10 \pm 0.003$, $p<10^{-16}$) with an average performance increment of $\Delta R=0.03$,
and DRE ($0.20 \pm 0.003$) outperforms DRE NO-INIT ($0.077 \pm 0.002\%$, $p<10^{-17}$) with an average performance increment of $\Delta R=0.12$. 

\paragraph{DRE's recurrence specifically captures alpha-dependent evoked responses.}
To further explore how DRE learns inter-trial variability, we investigate a well-known interaction between evoked responses and pre-stimulus activity. Specifically, brain responses to sensory input are known to be modulated by pre-stimulus oscillatory activity in the ``alpha" frequency range (8 - 13\,Hz) \citep{vanrullen2016perceptual}. To test whether this phenomenon can be detected in the present dataset, we compared the average evoked responses to words for ``high pre-stimulus alpha" versus ``low alpha" trials, using a median split for each subject separately. The results (Figure \ref{fig:alpha_effect}) show an effect of up to $50 \times 10^{-12}$T in fronto-temporal channels, peaking around 400\,ms after word onset. Critically, while the single-trial predictions of DRE capture this phenomenon, neither TRF nor RTRF learn to modulate their evoked responses depending on the alpha power (Figure \ref{fig:alpha_effect}B. Bottom).

This interaction between pre-stimulus alpha activity and evoked responses varies with the content of words, and more specifically, with their frequency in natural language: a factorial split between ``high alpha" \emph{versus} ``low alpha" trials and ``high word frequency" \emph{versus} ``low word frequency" trials resulted in both main and interaction effects (Figure \ref{fig:alpha_effect}C-D). Specifically, the comparison between these 2x2 conditions reveals three main phases. First, a main effect of alpha can be observed before the evoked response (light vs. dark lines in Figure \ref{fig:alpha_effect}C, $p < 10^{-3}$). Second, the main effect of word frequency starts to become significant from $\approx 200$\,ms (blue vs. red lines, $p < 10^{-4}$). Finally, the main effect of alpha starts to fade away after $\approx 500$\,ms ($p > 10^{-2}$ for high-frequency words), but its interaction with the stimulus continues to be significant ($p < 10^{-2}$). Critically, DRE learns these interactions between pre-stimulus alpha power and stimulus responses, while the linear models do not.

\paragraph{Feature importance helps interpreting the links between brain responses and stimulus features}
\begin{figure}[ht]
\centering
\includegraphics[width=\columnwidth]{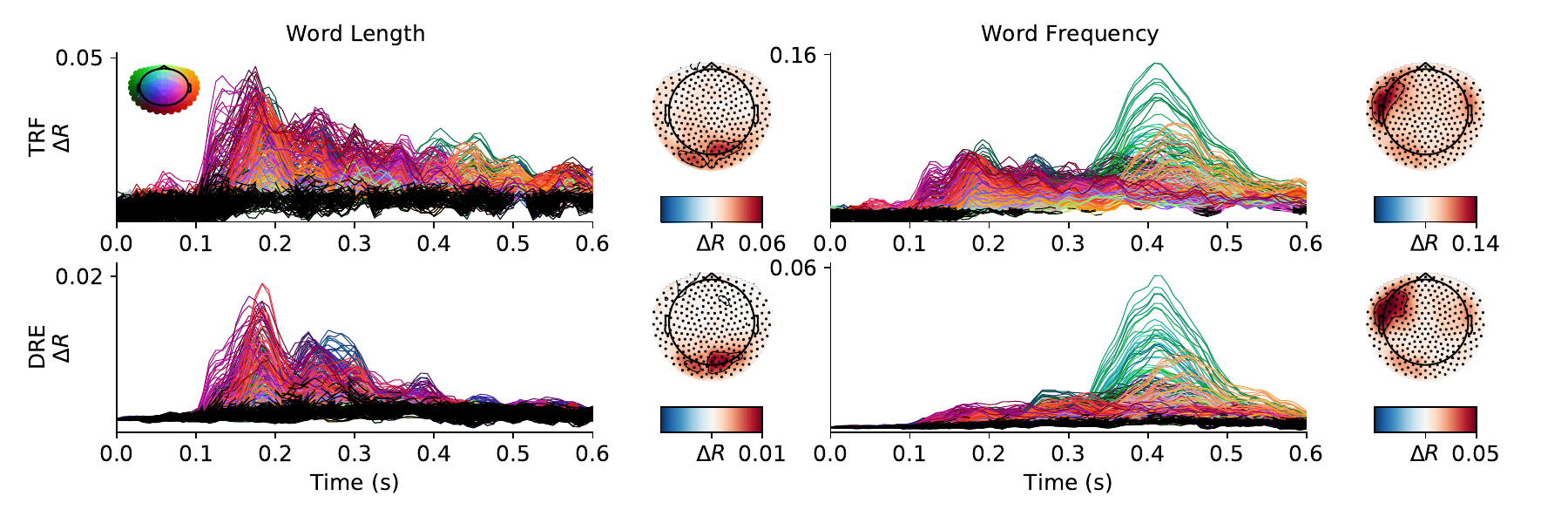}
\caption{Permutation importance ($\Delta R$) of word length (left column) and word frequency (right column), as a function of spatial location (color-coded by channel position, see top-left rainbow topography) and time relative to word onset for the two main encoding models (rows). The amplitudes of each line represent the mean across words and subjects for each channel. An FDR-corrected Wilcoxon test across subjects assesses the significance for each channel and time samples independently. Non-significant effects (p-value higher than 5\%) are displayed in black. 
Overall, this feature importance analysis confirms that early visual and late frontal responses are modulated by word length and word frequency, respectively, as expected.}
\label{fig:feature_importance}
\end{figure}

Interpreting nonlinear and/or high-dimensional models is notoriously challenging \citep{lipton2018mythos}. This issue poses strong limitations on the application of deep learning to neural recordings, where interpretability remains a major goal \citep{king2018encoding,ivanova2020simple}. While DRE faces the same types of issues as any deep neural network, we show below that a simple feature importance analysis of the predictions of this model (as opposed to its parameters) yields results that are consistent with those obtained by linear models, and with those described in neuroscientific literature (cf. Section ~\ref{ssec:model_evaluation}). 

Feature importance quantifies the loss of prediction performance $\Delta R$ when a unique feature is shuffled across words as compared to a non-shuffled prediction.  
Here, we focus our analysis on word length and word frequency, as these two features have been repeatedly associated with early sensory responses in the visual cortex and late lexical responses in the temporal cortex, respectively  \citep{kutas_n400,fedorenko2016neural}. 
As expected, the feature importance for word length in Figure \ref{fig:feature_importance} peaked around 150\,ms in posterior MEG channels, whereas the feature importance of word frequency peaked around 400\,ms in fronto-temporal MEG channels, for both the TRF and the DRE models. Furthermore, we recover a second phenomenon known in the literature: the \textit{lateralization} of lexical processing in the brain. Indeed, Figure \ref{fig:feature_importance} shows, for the word frequency, an effect similar in shape across hemispheres, but significantly higher in amplitude for the left hemisphere (e.g. $p<10^{-10}$ in the frontal region, $p<10^{-12}$ in the temporal region, for the DRE).

These results suggest that, in spite of being high dimensional and nonlinear, DRE can be interpreted similarly to linear models in the present context.

\section{Discussion}

The present study demonstrates that DRE outperforms several of its linear counterparts to predict MEG time series. In particular, it addresses the three challenges introduced above. First, the complex, nonlinear and non-stationary dynamics can be efficiently modeled by deep convolutional LSTM layers. Second, inter-trial and inter-individual variability can be addressed with recurrence (i.e. MEG-INIT) and subject embeddings, respectively. Finally, the relationship between stimulus features and brain responses can be interpreted in light of a permutation-based feature importance analyses.

Overall, the present study shows that the gain in prediction performance obtained by deep learning algorithms may not necessarily come at the price of interpretability. Indeed, we show here that DRE can be probed \emph{a posteriori} to reveal how evoked responses relate to each stimulus feature and/or to pre-stimulus brain activity. 
This feature importance supplements ongoing efforts to open black-box models of brain activity. For example, \citet{vangerven2017fmri} used a recurrent neural network to predict fMRI recordings, and quantified the impact of stimulus features by correlating them with the model's hidden state. 
Similarly, \citet{mesgarani2019}, analyzed the activations of a deep convolutional network with TRF to show how they captured ``dynamical receptive fields". In both of these cases, however, these post-hoc analyses are based on (i) linear assumptions and (ii) the inner  activations of the model. By contrast, the permutation feature importance used here focuses on probabilistic dependencies between input features and the models' predictions, which generalize linear dependencies measured by correlation \citep{Breiman}. This approach can thus be applied to any black-box predictive model.

Deep neural networks have not yet emerged as the go-to method for neuroimaging \citep{he2020deep, banvilledeepeeg2019}. Nevertheless, several studies have successfully used deep learning architectures to model brain signals \citep{richards2019deep, VanGerven2015, Aru2018,cadena2019cnn,Kording2018}. In particular, deep nets trained on natural images \citep{eickenberg2017seeing, bethge2017newcnn, yamins2021contrastive}, sounds \citep{mesgarani2019}, or text \citep{huth2018rnnlanguage} are being increasingly used as an \textit{encoding} model for predicting neural responses to sensory stimulation
\citep{yamins2016using, kriegeskorte2019neural,caucheteux2022brains,
millet2021inductive}.
Conversely, deep nets have also been trained to \textit{decode} sensory-motor signals from neural activity, successfully reconstructing text \citep{sun2019brain2char} from Electrocorticography (ECoG) recordings, or images \citep{guccluturk2017deep} from functional magnetic resonance imaging (fMRI). Despite these successes, we would like to argue that what possibly limits a wider impact of deep learning in human cognitive neuroscience is a combination of factors including: (i) low signal-to-noise ratio, (ii) small datasets, and (iii) a lack of high temporal resolution, where nonlinear dynamics may be the most prominent. The present experimental results make a step in this direction, and could thus open an avenue towards leveraging the many existing shorter naturalistic stimulus datasets collected on many subjects. This could be an alternative to making new long recordings of many hours of data from a handful of subjects \citep{ibc2018}.

While the DRE's architecture may be efficient at handling the dynamical structure of brain data, the dynamics assessed in this study are driven by specific linguistic features (i.e. word-length and word frequency). By contrast, recent ECoG and MEG studies have used more complex word features, represented as activations of a deep network pretrained on visual or language tasks \citep{caucheteux2020nlpbrain,schrimpf2020deepbrain,toneva2019nlpbrain}, and then predict the brain response in a way that is unaware of the dynamics (using a linear classifier for each time sample independently). Given the successes independently observed with these two approaches, a natural extension of this work would be to combine the two and learn to map complex stimuli to brain responses using (1) rich representations for the stimuli (such as the activations of a pretrained deep network), followed by (2) a rich dynamical model such as the DRE.
It is however worth pointing out that this approach would naturally lead to more high-dimensional parameter spaces, which would require larger datasets to limit potential overfitting.

The present work is based on a deterministic and predictive framework using deep learning. Other complementary approaches such as Hidden Markov Models (HMMs) and Gaussian Processes (GPs) have also been proposed to model brain data in a probabilistic framework. Such approaches have
been exploited to explore the spatio-temporal statistics of fMRI or MEG data~\citep{ambrogioni2017gp,vidaurre2016}, but also in an encoding context~\citep{vangerven2017,schoemakers2013gp}.
In particular, \cite{ambrogioni2017gp} combine GPs and dynamical system modeling to account for MEG responses to tactile input and shows that it captures meaningful modulations of oscillatory activity. This approach may offer a promising avenue to further clarify the interaction between baseline alpha oscillations and visual responses captured by DRE (see Figure~\ref{fig:alpha_effect}).
Similarly, \cite{schoemakers2013gp} show that Gaussian modeling efficiently learns to predict fMRI responses to visual stimuli, and, importantly, can be inverted to achieve zero-shot decoding of individual characters. By contrast, our encoding approach would necessitate additional fine-tuning to transfer DRE to a novel decoding task.
Overall, a key advantage of probabilistic models like GP is the ability to quantify uncertainty in the predictions, which in the present forecasting scenario would likely increase when looking at late latencies. While the
proposed approach does not offer this possibility, the present study benefits from the highly-optimized ecosystem of deep learning, which allows us to efficiently deal with the large size of raw MEG data (273 MEG channels sampled at 1,200\,Hz and recorded for 60 min in 99 subjects).

It is worth noting that because the losses used to train the models in this paper are MSEs evaluated in the time domain, the TRF, RTRF, and DRE are solely trained and evaluated on their ability to predict the \emph{amplitude} of the neural signal \emph{at each time sample}. Consequently, the objective solely focuses on ``evoked" activity, i.e. neural signals that are strictly phase-locked to stimulus onset or to past brain signals \citep{tallon1999oscillatory}. A quick time-frequency decomposition of the models suggest that none of them capture ``induced" activity, e.g. changes in the amplitude of neural oscillations with a non-deterministic phase. A fundamentally distinct loss would be necessary to capture such non phase-locked neural dynamics.

As for many other scientific disciplines, deep neural networks will undoubtedly complement -- if not shift -- the myriad of analytical pipelines developed over the years toward standardized end-to-end modeling. While such methodological development may improve our ability to predict how the brain responds to various exogenous stimuli, the present attempt already highlights the many challenges that this approach entails. Nevertheless, the present results hopefully clearly demonstrate that deep networks are a very relevant technology to capture complex neural dynamics collected non-invasively by MEG and certainly EEG.

\section*{Acknowledgements}

Experiments on MEG data were made possible thanks to MNE~\citep{MNE,MNEPython}, as well as the scientific Python ecosystem: Matplotlib~\citep{hunter2007matplotlib}, Scikit-learn~\citep{scikit-learn}, Numpy~\citep{harris2020array}, Scipy~\citep{2020SciPy-NMeth} and PyTorch~\citep{pytorch}.

This work was supported by the French ANR-20-CHIA-0016 and the European Research Council Starting Grant SLAB ERC-StG-676943 to AG, and by the French ANR-17-EURE-0017 and the Fyssen Foundation to JRK for his work at PSL.

The authors would like to thank Nicholas Tolley and Hubert Banville for their feedback and suggestions on the early versions of this manuscript.

\section*{Conflict of interest}
The authors declare no conflict of interest.

\bibliography{paper}



\newpage 

\section{Appendix}
\label{sec:appendix}

\begin{figure}[H]
\centering
\includegraphics[width=\columnwidth]{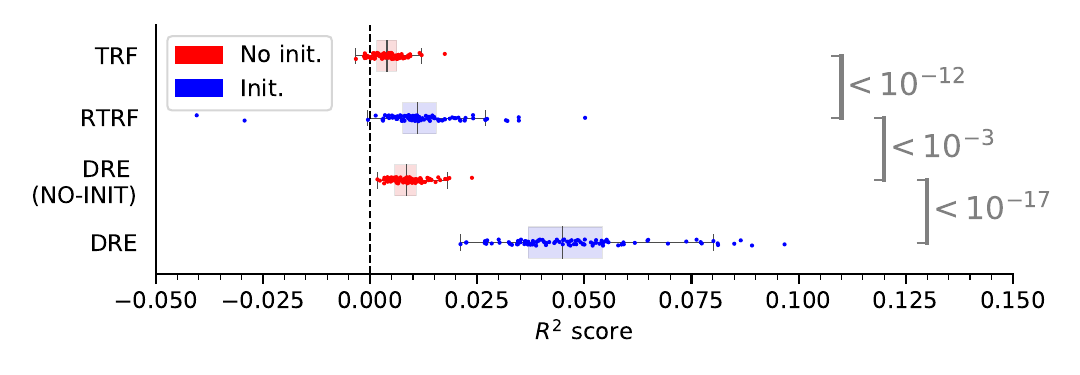}
\caption{Explained Variance captured by coefficient of determination ($R^2$) for each model. Boxplots indicate median and interquartile range, while each point corresponds to a single subject. The gray brackets indicate the p-value obtained with a pairwise Wilcoxon comparison across subjects. Initialization and non-linearity increase predictive performance.}
\label{fig:explainedvar}
\end{figure}

Figure~\ref{fig:explainedvar} reports an alternative scoring of the main models using the $R^2$ score instead of the Pearson correlation $R$. It is interpretable as explained variance is a quantity upper bounded by 1, as opposed to  MSE which depends on the input scaling of the data.
Due to the very small SNR of single trial and unaveraged MEG recordings, we obtain low values as expected. Indeed, some variance in the signal is explained by noise only, whose amplitude is on a scale 10 times larger that of the evoked response. However, the ordering of model performance and the conclusions are left unchanged.

\paragraph{Note on convolution and computational efficiency}
The introduction of convolutional layers is here mainly motivated by computational efficiency: convolutional layers reduced training time on an NVIDIA V100 GPU from 2.6\,h for a DRE devoid of convolutional layers down to 1.4\,h for our DRE. The performance between these two models is relatively similar, although with a slight benefit in favour of DRE:(NO-CONV: $0.194 \pm 0.003\%$; DRE: $0.197 \pm 0.003\%$, $p<10^{-5}$).

\newpage 

\begin{figure}[H]
\centering
\includegraphics[width=\columnwidth]{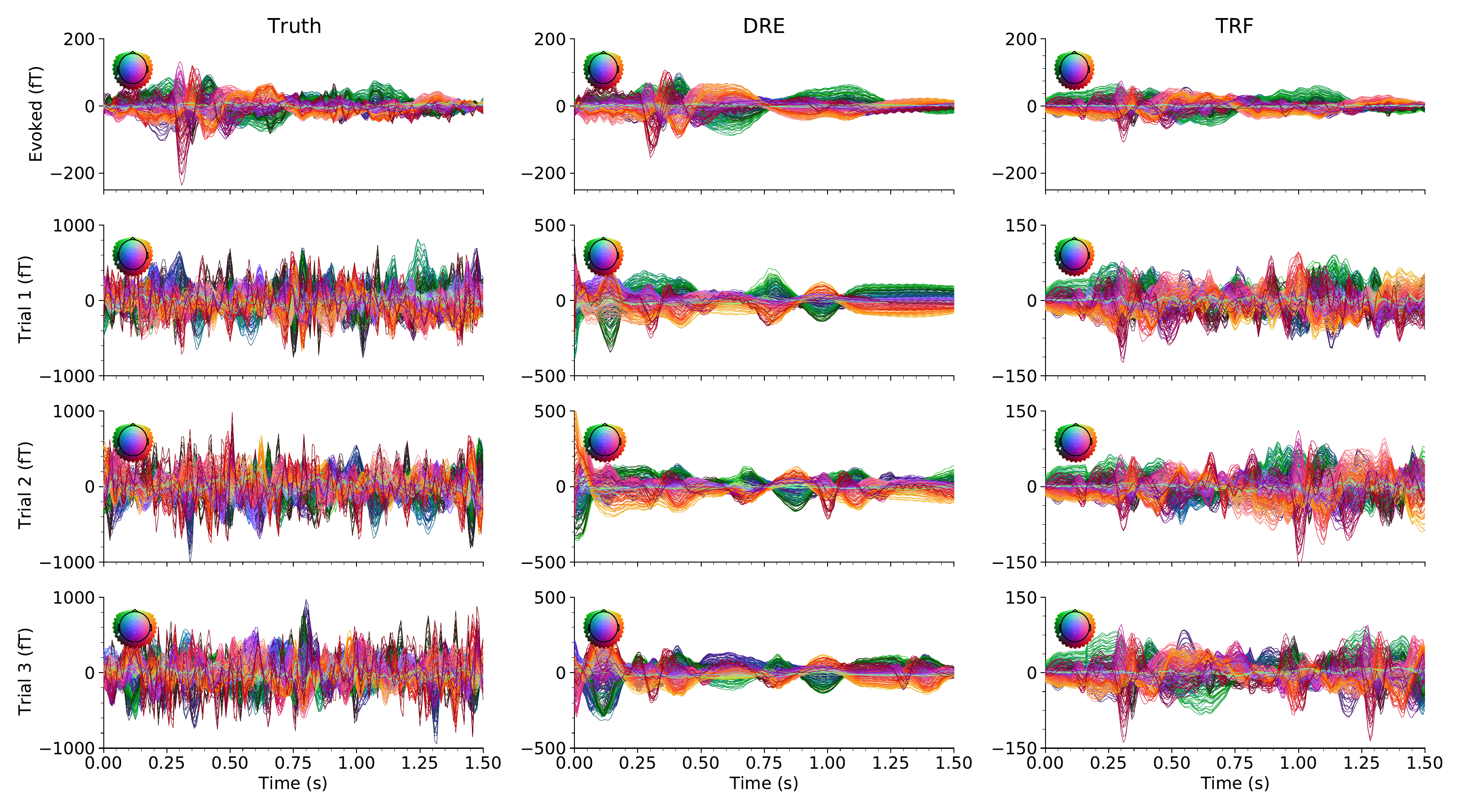}
\caption{True and predicted brain response (averaged ``Evoked" (top row) or single-trial) to visual stimuli.
These qualitative results illustrate that individual MEG responses vary along many different aspects, including the amplitude, latency and overall shape of the evoked-related fields. These variations are particularly challenging for TRF to learn. The predictions of TRF illustrated here also highlight that this model appears to modulate the MEG response but typically outputs predictions with reduced amplitudes -- an effect likely due to a reduction-to-the-mean phenomenon: \emph{i.e.} when the model fails to learn the MEG dynamics, it converges its coefficients towards 0.}
\label{fig:intertrialvariability}
\end{figure}

\end{document}